\documentclass[aps,prl,superscriptaddress,twocolumn]{revtex4}
\usepackage{bm} \usepackage{graphicx} \usepackage{amsmath}
\usepackage{amssymb} \usepackage{epstopdf} \usepackage{txfonts}

\begin{document}

\title{Non-Perturbative
Entangling Gates between Distant Qubits using Uniform Cold Atom Chains}

\author{Leonardo Banchi}
\affiliation{Dipartimento di Fisica, Universit\`a di Firenze,
Via G. Sansone 1, I-50019 Sesto Fiorentino (FI), Italy}
\affiliation{INFN Sezione di Firenze, via G.Sansone 1, I-50019 Sesto
Fiorentino (FI), Italy}

\author{Abolfazl Bayat}
\affiliation{Department of Physics and Astronomy, University College
London, Gower St., London WC1E 6BT, United Kingdom}

\author{Paola Verrucchi}
\affiliation{ISC - Consiglio Nazionale delle
Ricerche, UoS via G.Sansone 1, I-50019 Sesto Fiorentino (FI), Italy}
\affiliation{INFN Sezione di Firenze, via G.Sansone 1, I-50019 Sesto
Fiorentino (FI), Italy}
\affiliation{Dipartimento di Fisica, Universit\`a di Firenze,
Via G. Sansone 1, I-50019 Sesto Fiorentino (FI), Italy}

\author{Sougato Bose}
\affiliation{Department of Physics and Astronomy, University College
London, Gower St., London WC1E 6BT, United Kingdom}

\date{\today}

\begin{abstract}
We propose a new fast scalable method for achieving a two-qubit entangling
gate between arbitrary distant qubits in a network
by exploiting dispersionless propagation in uniform chains.
This is achieved dynamically by switching on a strong interaction
between the qubits and a bus formed by a non-engineered chain
of interacting qubits. The quality of the gate scales very efficiently
with qubit separations.
Surprisingly, a sudden switching of the coupling is not necessary and
our gate mechanism is not altered by a possibly gradual switching.
The bus is also naturally reset to its initial state making
the complex resetting procedure unnecessary after each application of
the gate.
Moreover, we propose a possible experimental realization
in cold atoms trapped in optical
lattices and near field Fresnel trapping potentials, which are both
accessible to current technology.
\end{abstract}

\maketitle

{\em Introduction:--}
Universal quantum computation can be achieved by arbitrary local
operations on single qubit and one two-qubit entangling gate
\cite{Bremner2002}.
While single qubit operations are easily achieved by local actions,
the story is very different for the two qubit gate.
In an array of spins an entangling gate between neighboring qubits can be
accomplished by letting them interact. However, for non-neighboring qubits,
a direct interaction is normally not possible unless there is a separate common bus mode
\cite{Cirac1995} or flying qubits.
In realizations without an additional bus mode,
such as with cold atoms in optical lattices,
one cannot  choose an arbitrary pair of atomic qubits for a gate operation
and usually gates parallely occur between all neighboring pairs
\cite{Jaksch1999}.
Thus, designing bus modes for logic gates between arbitrary and distant pairs
of qubits is of utmost importance in any physical realizations
and various unconventional examples of buses  are continuously being
proposed \cite{Stannigel2010Burkard2006,Yao2010}.
One possible realization is to have both the qubits and the bus composed
of the same physical objects, generally called spin chains.
The quality of an unmodulated spin chain, even as a data-bus,
is affected by dispersion \cite{Bose2003Bayat2008a}.
Thus, in order to have a quantum gate
between two qubits through such buses
\cite{Yung2005,Clark2005,gorshkov2010PRL,Yao2010},
delocalized encodings over several spins
\cite{OsborneL2004}, delicately engineered couplings \cite{Albanese2004} or
very weak couplings between qubits and the bus
\cite{Yao2010} is thought to be necessary.
Recently, a new scheme based on tuning the couplings between qubits
and the bus has been proposed \cite{Leonardo10} for \emph{fast} and
\emph{high-quality} state transmission, which we here exploit for
achieving an entangling quantum gate between arbitrarily distant qubits.

Cold atoms in optical lattices are now an established field for
testing many-body physics
\cite{Bakr2009,Brennen1999,Mandel2003,Greiner2002,Sherson2010Bakr2010}.
In particular,
chains of atoms in Mott insulator regime  (one atom per site) are being
built experimentally
\cite{Greiner2002,Sherson2010Bakr2010}, paving the way for realizing
spin Hamiltonians \cite{Lukin2003}.
With recent cooling methods, the required temperatures for
observing magnetic quantum phases has become
reachable \cite{Medley2010}. In this framework,
series of multiple two-qubit gates, acting globally and
simultaneously, have been
proposed \cite{Briegel2000} and realized \cite{Mandel2003}.
Could the same framework solve the problem of
realizing quantum gates between any two selected neutral
atom qubits? This is still an outstanding problem,
unless one uses the physical movements of neutral atoms
to each other's proximity \cite{cirac2000} which has its own complexity.
%While such movements may still have the ability to create a scalable
%neutral atomic quantum computer, alternative methodologies,
%without the complexity of mechanical processes,
%are worth pursuing for long range scalable gates.

Recently, single site addressing in an optical lattice setup
has been experimentally achieved \cite{Sherson2010Bakr2010}.
Furthermore, local traps have been proposed for individual atoms
using Near Field Fresnel Diffraction (NFFD) light \cite{Bandi2008}.
A new approach for scalable quantum computation
has been suggested \cite{Nakahara2010} through a combination
of local NFFD traps, for qubits, and an empty optical lattice,
for mediating interaction between them. Since the interaction
is achieved through controlled collisions
between \emph{delocalized} atoms it may suffer a high decoherence
when qubits, on which the gate is applied, are far apart \cite{Mandel2003}.

In this letter we put forward a scalable, non-perturbative
(i.e. not relying on weak couplings)
dynamical scheme for achieving high-quality
entangling gates between two arbitrarily distant qubits, suitable
for subsequent uses without resetting.
Unlike previous proposals, we do not demand encoding, engineering or
weak couplings: we only need switchable couplings between
qubits and  the bus. We also propose an application, based on a
combination of NFFD traps and optical lattices, which is
robust against possible imperfections.

{\em Introducing the model:--}
Let us describe our bus as a chain of spin $1/2$ particles interacting
through %the Hamiltonian
\begin{equation}\label{hb}
H_M = J \: \sum_{n=1}^{N-1} \left(\sigma_n^x \sigma_{n+1}^x +
\sigma_n^y \sigma_{n+1}^y + \lambda \:\sigma_n^z \sigma_{n+1}^z\right),
\end{equation}
where $\sigma_n^{\alpha}$ ($\alpha=x,y,z$) are Pauli operators acting on
site $n$, $J$ is the exchange energy and $\lambda$ is the anisotropy.
The qubits $A$ and $B$, on which the gate acts,
sit at the opposite sides of the bus, labeled by site $0$ and $N+1$
respectively.
The interaction between the bus and the qubits is %has the form of
\begin{equation}\label{hi}
H_I = J_0 \: \sum_{n=0,N} \left(\sigma_n^x \sigma_{n+1}^x +
		\sigma_n^y \sigma_{n+1}^y + \lambda\:
		\sigma_n^z \sigma_{n+1}^z \right),
\end{equation}
where the coupling $J_0$ can be switched on/off.
For the moment the anisotropy $\lambda$ is set to zero. Initially the qubits
are prepared in the states $|\psi_A\rangle$ and $|\psi_B\rangle$ and
decoupled from the bus which is in the state $|\psi_M\rangle$,
an eigenstate of $H_M$, for instance the ground state.
Since $H_M$ commutes with the parity operator $\prod_{n=1}^N (-\sigma_n^z)$
and with the mirror inversion operator, the state $|\psi_M\rangle$ has a definite parity $(-1)^p$,
for some integer $p$, and is mirror symmetric.
At time $t=0$ the coupling $J_0$
is switched on and the whole system evolves under the effect of
the total Hamiltonian $H=H_M+H_I$, i.e.
$|\Psi(t)\rangle=e^{-i H t}|\psi_A\rangle|\psi_M\rangle|\psi_B\rangle$.
In Ref.~\cite{Leonardo10} it was shown that by tuning $J_0$
to an optimal non-perturbative value $J_0^{\text{opt}}\simeq 1.05 J  N^{-1/6}$
the mirror-inversion condition
\cite{Yung2005} is nearly satisfied resulting in a fast high-quality transmission.
In fact, when $|\psi_B\rangle$ is initialized in either
$|0\rangle \equiv |{\uparrow}\rangle$ or
$|1\rangle \equiv |{\downarrow}\rangle$
an arbitrary quantum state of $A$ is transmitted almost
perfectly to $B$ after time $t^* \simeq (0.25 N + 0.52 N^{1/3})/J$.

The Hamiltonian $H$ is mapped to a free fermionic model
by Jordan-Wigner transformation
$c_n = \Pi_{k=0}^{n-1}(-\sigma_k^z)\sigma_n^-$ (where
$\sigma_n^\pm = (\sigma_n^x\pm\sigma_n^y)/{2}$) followed by a unitary
transformation $d_k = \sum_n g_{kn} c_n$. The total Hamiltonian finally
reads $H=\sum_k\omega_k \: d_k^\dagger d_k$ where the explicit form
of $g_{kn}$ and $\omega_k$ are given in
\cite{WojcikLKGGB2005,CamposVenutiGIZ2007}.
The dynamics in the Heisenberg picture is given by
$c_n(t) = \sum_m U_{nm}(t) c_m$ where $U_{nm}(t)=\sum_{k} g_{kn}g_{km}
e^{-i\omega_k t}$.
When the perfect transmission condition, i.e.  $J_0=J_0^\text{opt}$, is
satisfied we have $|U_{0,N+1}(t^*)|^2 \simeq 1$,
and thus we set $U_{0,N+1}(t^*) = e^{i\alpha_{_N}}$.

Notice that in any transmission problem there always might be an overall phase
which is irrelevant to the quality of transmission. However, exploiting
this phase is the heart of our proposal for obtaining an entangling
two-qubit gate between $A$ and $B$.
We define $|\Psi_{ab}\rangle=|\Psi(0)\rangle$
with $|\psi_{A}\rangle = |a\rangle$ and $|\psi_B\rangle=|b\rangle$
where $a,b=0,1$.
When $J_0$ is switched on the whole system evolves and at $t=t^*$ the states
of $A$ and $B$ are swapped, while the bus takes its initial state
$|\psi_M\rangle$, as a result of the mirror inverting dynamics.
Therefore, an almost perfect transmission is achieved
with an overall phase $\phi_{ab}$,
namely $ e^{-i Ht^*}|\Psi_{ab}\rangle\approx
e^{i \phi_{ab}}|\Psi_{ba}\rangle$.
The explicit form of $\phi_{ab}$ follows from the dynamics depicted above
with the freedom of setting $\phi_{00} = 0$. For instance
to get $\phi_{10}$ we have
\begin{align}
e^{-iHt^*}|\Psi_{10}\rangle = & e^{-iHt^*}c_0\:|\Psi_{00}\rangle\simeq
\: U_{0,N+1}(-t^*) \:c_{N+1}\: |\Psi_{00}\rangle =
\cr = & (-1)^{p+1} e^{-i\alpha_{_N}}\:|\Psi_{01}\rangle \equiv
e^{i \phi_{10}} \:|\Psi_{01}\rangle.
\end{align}
This defines $\phi_{10}=(p+1)\pi-\alpha_{_N}$ while
$\phi_{01}=\phi_{10}$ due to the symmetry of the system.
With similar argument we get $\phi_{11}=\pi-2\alpha_{_N}$.
Therefore, the ideal mirror-inverting dynamics defines a quantum
gate $G$ between $A$ and $B$, which reads
$G|ab\rangle=e^{i\phi_{ab}} |ba\rangle$ in the computational basis.
Independent of the value of $\alpha_{_N}$ when the pair $A,B$
is initially in the state of $|{+}{+}\rangle$,
where $|{+}\rangle=(|0\rangle+|1\rangle)/\sqrt2$,
the application of the gate $G$
results in a maximally entangled state between $A$ and $B$.
%Concurrence $C=|\sin(\phi_{11}/2 - \phi_{01})|$
Furthermore, the phase $\alpha_{_N}$ is found to be equal to
$\frac{\pi}2(N+1)$.

Since the dynamics is not perfectly dispersionless,
$|U_{0,N+1}(t^*)|$ is not exactly 1,
gate $G$ is not a perfect unitary operator.
In fact,
the dynamics of the qubits is described by a completely
positive map,
$\rho_{_{0,N+1}}(t)=\mathcal{E}_t\big[\rho_{_{0,N+1}}(0)\big]$,
which can be written in components as
$ \langle i|\rho_{_{0,N+1}}(t)|j\rangle=\sum_{k,l}
\mathcal{E}_{ij,kl}(t)\langle k|\rho_{_{0,N+1}}(0)|l\rangle$.
To quantify the quality of the gate we calculate average gate fidelity
$\mathcal{F}_G(t) = \int d\psi\,\langle\psi|G^\dagger\mathcal{E}_t\big[
|\psi\rangle\langle\psi|\big]G|\psi\rangle$ where the integration is over
all possible two-qubit pure states. Using the results of
\cite{Zanardi2004,Nielsen2002} we get
\begin{equation}\label{agf}
\mathcal{F}_G(t)=\frac{\sum_{i,j,k,l}G_{ik}^* \mathcal{E}_{ij,kl}(t)
	G_{jl} + 4}{20},
\end{equation}
where
$\mathcal{E}_{ij,kl}(t)=\langle i|\mathcal{E}_t\big[|k\rangle\langle l|\big]
|j\rangle$ are numerically evaluated.

\begin{figure}
\begin{center}\includegraphics[width=8.5cm,angle=0]{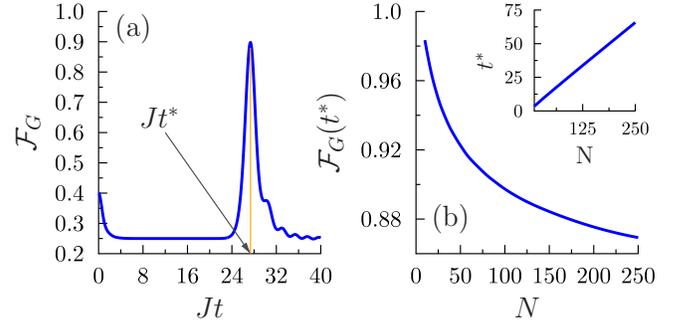}\end{center}
\caption{(Color online) (a) Evolution of the average gate fidelity
for a chain of $N=100$ and $J_0=0.5J$.
(b) $\mathcal{F}_G(t^*)$ as a function of $N$. Insets shows the optimal time
versus $N$.}
\label{fig1}
\end{figure}

\begin{figure}
\begin{center}\includegraphics[width=8.5cm,angle=0]{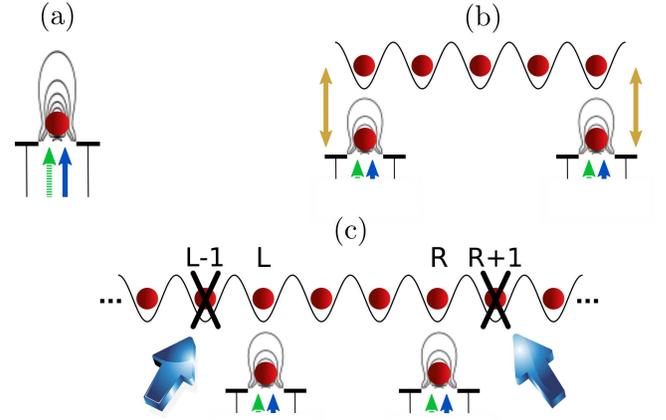}\end{center}
\caption{(Color online) (a) Local NFFD trap with two optical fibers,
one for trapping (solid blue) and one for unitary single qubit operations
(dashed green).  (b) Schematic interaction between qubits
(local traps) and the ending sites of the bus (optical lattice).
(c) Adiabatic cutting of the bus into three parts. }
\label{fig2}
\end{figure}

In Fig.~\ref{fig1}(a) we plot the time evolution of the average gate fidelity
for a bus of length $N=100$ initially in its ground state: $\mathcal{F}_G(t)$
displays a marked peak at $t=t^*$.
To show the scaling of the gate fidelity we plot
$\mathcal{F}_G(t^*)$ as a function of $N$ in Fig.~\ref{fig1}(b)
where we remarkably see that $\mathcal{F}_G(t^*)$ exceed 0.9
even for chains up to $N=100$ and decays very slowly with $N$.
Moreover, as shown in the inset of Fig.~\ref{fig1}(b) and unlike the
perturbative
schemes proposed in \cite{WojcikLKGGB2005,Yao2010}
our dynamics is fast.

Our dynamical gate works properly for arbitrary initial states of the bus
with fixed parity. Ideally after each gate application the parity of the
bus remains unchanged making it perfect for reusing. However, initialization in an eigenstate of $H_M$, besides
automatically fixing the parity, has the
advantage of simplicity for preparation.
Let us initially set the bus in its ground state and define
$F_M(t)$  as the fidelity between the ground state of $H_M$ and the
density matrix of the bus at time $t$.
To see how the quality of the gate operation is affected by $k$
subsequent uses of the bus, we compute
$\mathcal{F}_G(k t^*)$ and $F_M(k t^*)$ which are shown
in TABLE~I for $k=1,\dots,8$ subsequent uses.
\begin{table}
\begin{centering}
\begin{tabular}{|c|c|c|c|c|c|c|c|c|}
\hline
$k$ &1 & 2 & 3 & 4 & 5 & 6 & 7 & 8 \\
\hline
$\mathcal{F}_G(k t^*)$  & 0.984 & 0.961& 0.939 & 0.918 &0.898 & 0.879& 0.861& 0.844\\
\hline
$F_M(k t^*)$  & 0.966 & 0.926& 0.884 & 0.840& 0.795& 0.748& 0.701& 0.654\\
\hline
\end{tabular}
\label{table_1}
\caption{$\mathcal{F}_G(k t^*)$ and $F_M(k t^*)$ for up to 8 subsequent
uses of the bus of length $N=8$ without resetting.  }
\par\end{centering}
%\centering{}
\end{table}

{\em Application:--}
We now propose an application of the above gate mechanism for a scalable
neutral atom quantum computer with qubits held in static traps.
We consider a network of qubits each encoded in
two degenerate hyperfine levels of a neutral atom, cooled and localized
in a separate NFFD trap \cite{Bandi2008}.
In Fig.~\ref{fig2}(a) we show a single atom confined in a NFFD trap.
The position of the minimum of the trapping potential is controlled
by varying the aperture radius \cite{Bandi2008} through
micro electro mechanical system technology, as proposed in \cite{Nakahara2010}.
Local unitary operation on each qubit may be applied through an
extra fiber, along with the NFFD trapping fiber \cite{Nakahara2010}, as
show in Fig.~\ref{fig2}(a).
The qubits in the network are connected by a bus realized
by cold atoms in an optical lattice, prepared in the Mott
insulator regime \cite{Greiner2002,Sherson2010Bakr2010}. The
polarization and intensity of lasers are tuned so that
one ends up \cite{Lukin2003} with an effective Hamiltonian of Eq.~\eqref{hb}.
For the moment we assume that the distance between the two qubits,
on which we want to apply the gate, is equal to the length of the lattice
such that the two qubits interact with the atoms in the ending sites of the
lattice, as shown in Fig.~\ref{fig2}(b). To
switch on the interaction $H_I$ between the qubits and the bus we have
to move the minimum of NFFD trapping potential slightly higher such that
the qubits move upwards and sit at a certain distance from
the ends of the lattice.
By controlling such distance one can tune the interaction coupling to be
$J_0^{\text{opt}}$.
In order to simultaneously obtain interactions effectively described by
$H_M$ and $H_I$ we have to use the same spin dependent trapping laser beams
in both NFFD traps and optical lattice.

\begin{figure}\begin{center}
\includegraphics[width=8.5cm,angle=0]{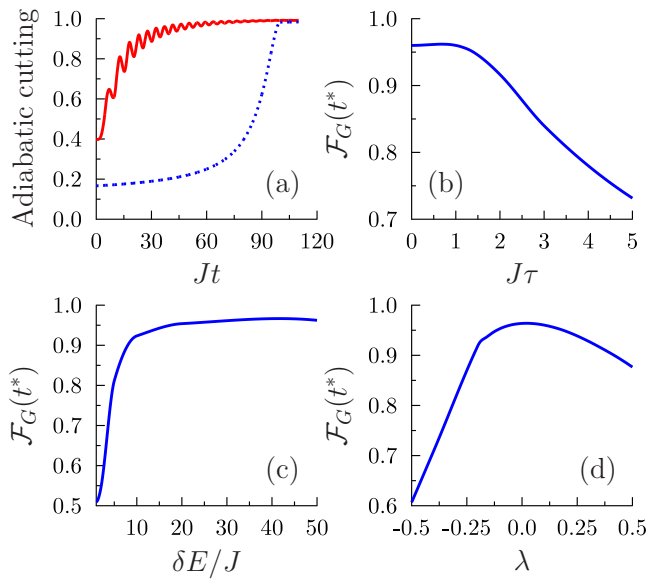}
\end{center}
\caption{(Color online) (a) Evolution of $F_M(t)$ under the adiabatic cutting
(solid red). Evolution of the fidelity between the ground state of the
whole optical lattice with the state of the split one under the adiabatic
gluing (dashed blue).
$\delta E /J $ is linearly varied between 0 and 30 over the time $100/J$.
(b) $\mathcal{F}_G(t^*)$  vs. switching time $\tau$
over which $J_0$ is linearly switched on from $0$ to $J_0^\text{opt}$.
(c) $\mathcal{F}_G(t^*)$ vs.
$\delta E/J$ after adiabatic cutting of the optical lattice.
(d) $\mathcal{F}_G(t^*)$ as a function of anisotropy $\lambda$.
The length of the bus is set to $N=16$.
}
\label{fig3}
\end{figure}

Now we consider the situation in which the optical lattice size
is larger than the distance between the qubits $A$ and $B$ (see Fig.~\ref{fig2}(c)). In this case if we simply switch
on the interaction between qubits and two intermediate sites ($L$, $R$)
of the optical lattice, shown in Fig.~\ref{fig2}(c),
the two external parts of the lattice
play the role of environment and deteriorate the quality of the gate.
To preserve the gate quality we need to cut the lattice into
three parts and separate the bus,
extended from $L$ to $R$, from the rest of the optical lattice.
This can be done by adiabatically shining a
localized laser beam on the atoms sitting on sites $L-1$ and $R+1$
to drive them off resonance, as shown in Fig.~\ref{fig2}(c).
In this case driving the atom effectively generates a Stark shift between
the two degenerate ground state through a highly detuned classical laser beam
with strength $\Omega$ and detuning $\Delta \gg \Omega$.
This provides an energy shift $\delta E = \Omega^2/\Delta$
between the two degenerate ground states, which can be treated as a local
magnetic field in the $z$ direction on sites $L-1$ and $R+1$. Keeping
$\Omega/\Delta$ small one can control the strength $\Omega$
and detuning $\Delta$ such that $\delta E$ becomes larger than $J$.
When $\delta E \gg J$ the bus is separated
from the external parts of the optical lattice. Moreover,
as $\delta E$ adiabatically increases, the bus moves
into its ground state, meanwhile splitting up from the rest.
Despite the gapless nature of Hamiltonian
\eqref{hb} there is always a gap $\propto J/N$ due to the finite
size of the bus which guarantee the success of the adiabatic evolution.
In Fig.~\ref{fig3}(a) we plot $F_M(t)$
over the course of adiabatic cutting
when the whole lattice is initially in its ground state.
In this adiabatic evolution
$\delta E$ is linearly increased from $0$ to $30 J$ over the time
interval of $100/J$.
Once the bus been prepared in
its ground state  the gate operation can be accomplished as discussed above.
After the operation of the gate one may want to glue  the previously
split optical lattice and bring it back into its ground state.
This can be done easily by adiabatically switching off
$\delta E$ as shown in  Fig.~\ref{fig3}(a) where the fidelity of the state
of the whole optical lattice with its ground state is plotted.

{\em Time scale:--}
We now give an estimation of $t^*$ in the {\it worst} case
scenario where $A$ and $B$ sit on the boundary of the lattice,
 which typically consists of $N \simeq 100$.
The typical values for $J$ in optical lattices are few hundred Hertz
(e.g. $J=360 \,h\,\text{Hz}$ in \cite{Trotzky2010}).
From the inset of Fig.~\ref{fig1}(b) we get $Jt^* \simeq 30$ for
$N=100$ and thus $t^* \simeq 13 \,\text{ms}$ which
is well below the typical decoherence time of the hyperfine
levels ($\simeq 10\,\text{minutes}$ \cite{bollinger1991303}).
Though there are some recent realizations of entangling gates \cite{Wilk2010Saffman2010}, faster
than ours, they are much less versatile as they design a single,
very specific, isolated gate and do not construct the
gate as part of an extended system.
Considering this latter kind
of architecture, our mechanism is much faster than
the perturbative methods \cite{Yao2010,WojcikLKGGB2005},
and operates at the time scale of $\mathcal{O}(N/J)$ which is
the best possible in any physical realization.

%Different schemes for realizing fast entangling gates have
%been recently put forward \cite{Wilk2010Saffman2010}, with
%operating times shorter than ours.
%However, such schemes are much less versatile as far as
%computational tasks are concerned, as they design a single,
%very specific, isolated gate and do not construct the
%gate as part of an extended system of connected qubits.
%On the other hand, when considering this latter kind
%of architecture, our mechanism is much faster than
%the perturbative methods \cite{Yao2010,WojcikLKGGB2005},
%and operates at the time scale of $\mathcal{O}(N/J)$ which is
%the best possible in any physical realization.

%Though, the mechanisms in which atoms are physically moving
%\cite{Wilk2010Saffman2010} are faster than ours,
%our proposal offers an alternative
%paradigm without involving physical movement.
%In fact, our mechanism is much faster than the perturbative methods
%\cite{Yao2010,WojcikLKGGB2005} for stationary qubits and operates at
%the time scale of $\mathcal{O}(N/J)$ which is
%the best possible in any physical realization.

{\em Imperfections:--}
Cold atom systems are usually clean and almost decoherence free;
however,
in the above proposed setup there might be some sources of destructive
effects which may deteriorate the quality of our scheme.
In particular, we consider: (i) gradual switching of $J_0$; (ii) imperfect
cutting of the chain when $\delta E$ is not large enough; (iii) existence
of interaction terms in the Hamiltonian which alter its
non-interacting free-fermionic nature.
In Fig.~\ref{fig3}(b) we show $\mathcal{F}_G(t^*)$ when $J_0$ is
gradually switched on from $0$ to $J_0^{\text{opt}}$
according to $J_0(t) = J_0^{\text{opt}} t/\tau$,
as a function of switching time $\tau$. It is indeed of general relevance
that a plateau
over which $\mathcal{F}_G(t^*)$ remains constant is observed, even for
$\tau$ as long as $1/J_0$.
In Fig.~\ref{fig3}(c) we plot
$\mathcal{F}_G(t^*)$ as a function of the energy splitting $\delta E$ on which
the cutting process is based. As it is clear from Fig.~\ref{fig3}(c), when
$\delta E > 10 J$ the bus is well isolated from the external parts,
which guarantees the high quality of the gate.
We have also
studied the effect of the anisotropy $\lambda$, possibly entering $H_M$ and
$H_I$, due to imperfect tuning of laser parameters \cite{Lukin2003}.
In Fig.~\ref{fig3}(d) we plot $\mathcal{F}_G(t^*)$ as a function of $\lambda$
and observe weak deterioration of the gate quality as far as
$|\lambda|<0.2$.

{\em Conclusions:--}
In this letter, we have proposed a scalable scheme for realizing a two-qubit
entangling gate between arbitrary distant qubits.
In our proposal, qubits are made of localized objects which
makes single qubit gates affordable.
The qubits interact dynamically via an
extended unmodulated bus which does not need being specifically
engineered and, besides embodying a quantum channel,
actively serves to operate the entangling gate.
Moreover, thanks to the non-perturbative interaction between the qubits
and the bus our dynamics is fast, which minimizes destructive decoherence
effects.
Provided the coupling between the qubits and the bus is properly tuned,
the dynamical evolution of the whole system is essentially dispersionless,
thus allowing several subsequent uses of the bus without resetting.
Surprisingly, a sudden switching of the coupling is not necessary and
our fast dynamical gate mechanism is not altered by a possibly gradual
switching: this is of absolute relevance, not only from practical
viewpoints but also in a theoretical perspective.
Our proposal is general and can be implemented in various physical
realizations. Specifically
we have proposed an application based on
neutral atom qubits in an array of separated NFFD traps connected
by an optical lattice spin chain data bus, which both
are accessible to the current technology.

{\em Acknowledgements:--} Discussion with A. Beige, S. C. Benjamin,
K. Bongs, J. Fitzsimons, D. Jaksch, D. Lucas, F. Renzoni,
and T. Tuffarelli are kindly acknowledged.
SB and AB acknowledge the EPSRC,
and the Royal Society and the Wolfson Foundation.
LB thanks UCL for the kind hospitality.

\end{document}